\documentclass[amsmath,amssymb,aps,pre,letterpaper,twocolumn,floatfix]{revtex4-1}
\usepackage{graphicx} 
\usepackage{bm}
\usepackage{epstopdf}
\usepackage{textcomp}
\usepackage{gensymb}
\usepackage{array,multirow}
\newcolumntype{K}[1]{>{\centering\arraybackslash}m{#1}}
\usepackage{makecell}
\usepackage{tablefootnote}
\usepackage{blindtext}

\newcommand{\gamhat}{\hat{\gamma}}

\graphicspath{{./figures/}}
\newcommand{\keq}{K_\text{eq}}
\newcommand{\fl}[1]{\widehat{\widetilde{#1}}}
\renewcommand{\(}{\left(}
\renewcommand{\)}{\right)}


\begin{document}
	
	\title{Multi-modal transport and dispersion of organelles in narrow tubular cells}
	
	\author{Saurabh Mogre}
	\affiliation{Department of Physics, University of California San Diego, La Jolla, California}
	\author{Elena F. Koslover}
	\email{ekoslover@ucsd.edu}
	\affiliation{Department of Physics, University of California San Diego, La Jolla, California}
	\date{\today}
	\preprint{}

\begin{abstract}
Intracellular components explore the cytoplasm via active motor-driven transport in conjunction with passive diffusion. We model the motion of organelles in narrow tubular cells using analytical techniques and numerical simulations to study the efficiency of different transport modes in achieving various cellular objectives. Our model describes length and time scales over which each transport mode dominates organelle motion, along with various metrics to quantify exploration of intracellular space.
 For organelles that search for a specific target, we obtain the average capture time for given transport parameters and show that diffusion and active motion contribute comparably to target capture in the biologically relevant regime. Because many organelles have been found to tether to microtubules when not engaged in active motion, we study the interplay between immobilization due to tethering and increased probability of active transport. We derive parameter-dependent conditions under which tethering enhances long range transport and improves the target capture time. These results shed light on the optimization of intracellular transport machinery and provide experimentally testable predictions for the effects of transport regulation mechanisms such as tethering.
\end{abstract}

\maketitle
\newpage

\section{Introduction}
\label{sec:Introduction}

Transport of cargo within the intracellular environment is a highly essential and tightly regulated process. Most eukaryotic cells have an active transport machinery consisting of molecular motors moving on a network of cytoskeletal polymers such as microtubules or actin filaments. Organelles can couple directly to motor proteins via specialized adaptors~\cite{fu2014integrated}, or hitch-hike on other motile organelles~\cite{salogiannis2016peroxisomes}. This mode of transport results in motion that is processive over variable length scales up to many microns. Many organelles execute bidirectional motion, switching direction between processive runs by either engaging alternate motor types or transferring to a cytoskeletal track with different orientation~\cite{kural2007tracking,ross2008cargo,mudrakola2009optically,balint2013correlative,hancock2014bidirectional}.

In addition to this motor-driven processive transport, effectively diffusive motion of organelles can arise due to thermal noise, active fluctuations of cytoskeletal networks~\cite{brangwynne2009intracellular}, or hydrodynamic entrainment in flow set up by moving motors and cargo~\cite{mussel2014drag}. Evidence has shown that the short time-scale movement of organelles appears effectively diffusive even when the underlying cytoplasmic medium is primarily elastic ~\cite{brangwynne2009intracellular,jaqaman2011cytoskeletal,ananthanarayanan2013dynein}. For brevity, we will refer to this stochastic motion of organelles as passive diffusion, while acknowledging that the fluctuations underlying the motion can have a number of actively driven origins.

The interplay between active and passive transport modes gives rise to length-scale dependent effects. While active transport is efficient at transporting cargo over relatively long cellular distances, diffusion can more effectively spread organelles over smaller length scales. Prior modeling work demonstrated the role of multi-modal transport in intracellular signaling and reduction of noise, particularly in the case where organelles cannot carry out their biological functions in the actively walking state~\cite{godec2015signal,benichou2011intermittent}. Other theoretical studies have demonstrated that limited processivity can enhance target search even in the absence of a passive or diffusing state, highlighting the importance of bidirectional motion~\cite{campos2015optimal}. These results suggest that the transport machinery in the cell may be optimized to allow substantial contributions from both processive and diffusive transport. Endosomes, peroxisomes, lipid droplets, mitochondria and mRNA are some example intracellular species known to employ multi-modal transport to move around within the cell~\cite{tanaka1998targeted
	,kural2005kinesin
	,schuster2011transient
,targett2003live}.

The organization of the cytoskeletal network has a potentially important role to play in the distribution of intracellular particles.
While a number of past models for intracellular transport employed a continuum approximation for cytoskeletal density~\cite{bressloff2013stochastic,godec2015signal,gou2014mathematical}, it is becoming clear that the specific arrangement of distinct cytoskeletal tracks has a substantial impact on cargo transport~\cite{ando2015cytoskeletal}.
Obstructions due to intersecting microtubules may cause particles to pause or switch tracks and change the direction of movement~\cite{balint2013correlative}. Localized traps arising from heterogeneous filament polarity have been found to hinder transport in cell-scale computational models~\cite{ando2015cytoskeletal}.
In tubular cell projections such as neuronal axons and fungal hyphae tips, the arrangement of cytoskeletal filaments is highly simplified, with microtubules aligned along the tubular axis and in many cases uniformly polarized towards the distal tip~\cite{goldstein2000microtubule,egan2012microtubule}. These projections range in length from tens to many hundreds of microns, and require cargo to be efficiently transported from the cell body to the distal tips and back again. In addition to being particularly amenable to theoretical models of transport phenomena, these cell types are of fundamental biological importance. 
 Defects in axonal transport in neurons have been implicated in a number of human pathologies, ranging from multiple sclerosis to Alzheimer's to prion diseases~\cite{liu2012pathologies} . Due to their simplified morphology and long length, these tubular cells form an ideal system for investigating the length-scale dependent effects of multimodal transport.


The discrete nature of cytoskeletal tracks within tubular cell projections limits active transport to narrow axially oriented bundles of microtubules~\cite{egan2012microtubule}. It has been proposed in several cellular systems that transport efficiency is increased by directly tethering organelles to the microtubules in order to prevent them from losing access to the tracks~\cite{hancock2014bidirectional,cooper2009diffusive,culver2006microtubule}.  Tethering can occur by specialized adaptor proteins binding the organelle to cytoskeletal tracks, as in the case for axonal mitochondria that become preferentially anchored in cellular regions with high metabolic needs~\cite{kang2008docking,pekkurnaz2014glucose,frederick2007moving}.
 Alternately, the binding of multiple motor proteins to individual vesicles results in a tethering effect that is believed to contribute to observed motor cooperativity~\cite{hancock2014bidirectional,muller2008tug}.
Because tethering is expected to hinder short-range dispersion while enhancing the ability of organelles to engage in long-range processive walks, it can potentially serve as a regulatory mechanism for length-dependent transport.

 A variety of cellular processes rely on efficient transport to achieve distinct objectives necessary for biological function. One such objective is the establishment of a uniform distribution of particles throughout the cell, as is observed for peroxisomes, mitochondria and lipid droplets~\cite{lin2016active,chang2006mitochondrial, valm2017applying}.
 Establishing this distribution, starting from the point of genesis of particular organelles, requires rapid transport and broad dispersion across long cellular length-scales.
Another objective is the delivery of organelles to specific subcellular regions. Examples include the motion of synaptic vesicles from the cell body to the pre-synaptic terminal of neuronal axons~\cite{hirokawa2005molecular}, and the transport of vesicles containing newly synthesized membrane-bound proteins from the Golgi apparatus to the cell boundary~\cite{de2008exiting}. The role of different transport modes in this process depends on the length scale of separation between the site of organelle synthesis and their eventual target. 
A third cellular objective is the rapid encounter between an intracellular target and any one of a uniformly distributed population of organelles. For instance, peroxisomes serve to neutralize oxidative metabolic byproducts, and the health of a cell is dependent on rapid removal of these toxic species as soon as they appear~\cite{singh1997biochemistry}. Similarly, early endosomes rely on contact with any of a population of lysosomes that aid in releasing the endosomal contents into the cytoplasm~\cite{bright2005endocytic}.
The efficiency of such target encounter depends both on the nature of transport processes for the organelles and on their density within the cell.

In this article, we present a simplified model for transport in a tube through a combination of processive walks and diffusion. We analyze the relative contributions of the two transport modes, as well as the possibility of tethering to cytoskeletal tracks, in achieving the different transport objectives of the cell.
  Sec.~\ref{sec:model} establishes our halting creeper model and its behavior in terms of the rate with which particles explore a one-dimensional environment.  
  In Sec.~\ref{sec:dispersion}, we use the developed model to study the effects of bidirectional transport on distributing particles uniformly within a domain. In Sec.~\ref{sec:targetsearch}, we explore the contribution of different transport modes to the delivery of individual particles, as well as target clearance by a dispersed particle population. 
 Sec.~\ref{sec:tethering} introduces an expanded model that accounts for particle tethering, delineating the effects of this mechanism on organelle dispersion and target capture times. 

\section{Halting creeper model}
\label{sec:model} 

\begin{figure}[b!]
	\centering
	\includegraphics[width=\textwidth/3]{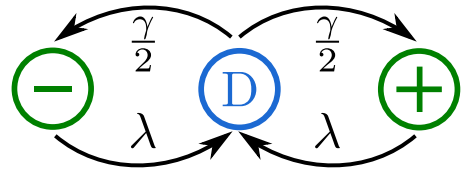}
	\caption{\label{fig:schematic} Schematic for the transition between particle states. `D' denotes diffusive particles, `+' denotes particles moving processively in the rightward direction, whereas `-' denotes particles moving processively in the leftward direction. The arrows are labeled with transition rates between states.}
\end{figure}

We define a simplified stochastic model for intracellular particles undergoing multi-modal transport, focusing on motion along a single dimension. 
Each ``halting creeper" particle exists in either a passive diffusive state characterized by diffusion coefficient $D$ or an actively moving state with constant speed $v$ in either the positive or negative direction (Fig.~\ref{fig:schematic}). Switching between the states is a Markovian process with constant starting rate $\gamma$ for transitioning from the passive to the active state and constant stopping rate $\lambda$ for transitioning from active to passive. Selection of the direction of motion is random at each initiation of an active run, and we assume complete symmetry between forward and backward motion. 
 We note that this model is a more general form of previously defined creeping particle models~\cite{campos2015optimal}, which have been analyzed in the limit $\gamma \rightarrow \infty, D \rightarrow 0$. Furthermore, a three-dimensional version of our halting creeper model has previously been explored in the context of mean squared displacement and local concentration fluctuations~\cite{godec2015signal}. By contrast, in this work we focus explicitly on the efficiency with which such two-state transport distributes organelles throughout a cell and delivers them to intracellular targets. 
 
 Two important quantities which describe the behavior of a halting creeper particle are the active run-length ($\ell = v/\lambda$)  and the equilibrium fraction of particles in the active state ($f = \frac{\gamma}{\lambda + \gamma}$).
 For much of the subsequent discussion, we non-dimensionalize all length units by the run-length $\ell$ and all time units by the run-time $1/\lambda$. We denote the remaining dimensionless parameters as $\hat{D} = \frac{D}{\ell^2\lambda}, \hat{\gamma} = \frac{\gamma}{\lambda}$, and  dimensionless time as $\hat{t} = \lambda t$.
 
 The Markovian nature of the transitions between active and passive states allows the calculation of a spatio-temporal propagator function $G(x,t)$ for the halting creeper, which gives the distribution of positions at time $t$ given the particle started at the origin at time $0$. This propagator is obtained by convolution in the space and time domain of the individual propagators for passive and active transport. After a Fourier transform in space ($\hat{x} \rightarrow k$, $G\rightarrow \widetilde{G}$) and a Laplace transform in time ($\hat{t} \rightarrow s$, $\widetilde{G}\rightarrow \widehat{\widetilde{G}}$), the multi-modal propagator for particles initially in an equilibrium distribution between passive and active states is given by
\begin{widetext}
\begin{equation}
\begin{split}
 \fl{G}(k,s) = \frac{(\lambda+s)(\gamma+\lambda)(\gamma+\lambda+s)+\left(D\gamma(\lambda+s)+v^2 \lambda \right)~k^2}{(\gamma+\lambda)\left[s(\lambda+s)(\gamma + \lambda +s)+\left(D(\lambda+s)^2+v^2(\gamma+s)\right)~k^2+ v^2 D ~k^4\right]},
\end{split}
\label{eqn:propagator}
\end{equation}
\end{widetext}
as derived in Appendix~\ref{app:propagator}. This propagator serves as the basis for our subsequent calculations on the efficiency of particle spreading and target site search.

\begin{table*}[t!]
	\centering
	\bgroup
	\def\arraystretch{1.5}
	\begin{tabular}{K{0.2\textwidth}K{0.13\textwidth}K{0.14\textwidth}K{0.1\textwidth}K{0.1\textwidth}K{0.12\textwidth}K{0.12\textwidth}K{0pt}}
		\hline
		\hline
		\multirow{3}{0.2\textwidth}{\centering Transport System} & Rate of switching to active transport  & Rate of switching to passive transport  & Velocity of active transport  & Diffusivity  & Density of the population & Approx. size of cellular region&\\
		& $\gamma~(s^{-1})$  & $\lambda~(s^{-1})$ & $v~(\mu\text{m}/s)$ & $D~(\mu\text{m}^2/s)$ & $\rho~(\mu\text{m}^{-1})$ & $L~(\mu\text{m})$ &\\[5pt]
		\hline
		Peroxisomes in fungal hyphae~\cite{lin2016active} & $0.015$\footnotemark[1] & $0.29$ & $1.9$ & $0.014$ & $1.5$ & $50$& \\
		Lysosomes in kidney cells~\cite{balint2013correlative,bandyopadhyay2014lysosome}
			 & $0.17$ & $0.15$ & $0.52$ & $0.071$ & -- & $20$&\\ 
		Mouse neuron transport vesicles in vitro~\cite{hendricks2010motor} & $0.33$\footnotemark[1] & $2.7$ & $0.8$ & $0.03$ & $0.14$ &--&\\
		Mitochondria in {\em Drosophila} axons~\cite{pilling2006kinesin} & $0.17$ & $0.15$ & $0.35$ & -- & $1.3$ &$1000$&\\
		Dense core vesicles in {\em Aplysia} neurons~\cite{ahmed2014active,romanova2006self} & $0.22$\footnotemark[1] & $2.2$\footnotemark[2]& $0.36$\footnotemark[3] & $0.002$\footnotemark[3] & $1.7$ & $100$ &\\
		PrP$^{\text{C}}$ vesicles in mouse axons~\cite{encalada2011stable,hendricks2010motor} & $0.36$ & $0.15$ & $0.85$ & $\bf{-}$ & $0.4$ & $100$&\\[-45pt]
		\multicolumn{8}{l}{\footnotetext[1]{Estimated from equilibrium fraction in active state.}}\\
		\multicolumn{8}{l}{\footnotetext[2]{Estimated from single particle trajectory.}}\\					
		\multicolumn{8}{l}{\footnotetext[3]{Estimated from MSD plot}}\\
		\hline
		\hline
	\end{tabular}
	\egroup
	\caption{Estimated values of transport parameters for some biological systems. Run length can be obtained as $\ell=v/\lambda$. Parameters can be converted to dimensionless units according to: $\hat{D} = D\lambda/v^2$, $\hat{\gamma} = \gamma/\lambda$, $\hat{\rho} = \rho v/\lambda$.}
	\label{tab:paramlist}
\end{table*}

\subsection{Particle spreading: mean squared displacement}
\label{subsec:msd}

 The mean squared displacement (MSD) is a commonly used measure of spreading speed for diffusing particles. For the halting creeper model, it can be calculated directly from the propagator as
\begin{equation*}
\begin{split}
\left<\hat{x}(\hat{t})^2\right> & = \mathcal{L}^{-1}\left[\left.-\frac{\partial^2 \widehat{\tilde{G}}}{\partial k^2}\right|_{k=0}\right] \\
& = 2 (1-f) \hat{D} \hat{t} + 2 f \left[\hat{t} + (e^{-\hat{t}}-1)\right],
\end{split}
\label{eqn:MSD}
\end{equation*}
where the Laplace inversion $\mathcal{L}^{-1}$ is carried out analytically via the residue theorem. 

This expression for the MSD is composed of a linear superposition of fraction $1-f$ of diffusing  particles and fraction $f$ of particles undergoing active walks that are persistent over a dimensionless time-scale of $1$. The latter component corresponds to an MSD that scales ballistically as $f\hat{t}^2$ for $\hat{t}\ll 1$ and diffusively as $2f\hat{t}$ for $\hat{t}\gg 1$. In the case of small diffusivity, there is an additional transition time when the ballistic motion begins to dominate over the passive diffusion. This occurs  at $\hat{t}^* = \frac{2\hat{D}}{\hat{\gamma}}$. When $\hat{t}^* \ll 1$, the MSD transitions from diffusive to ballistic and back to diffusive scaling (Fig.~\ref{fig:range}a). The long-time behavior of the particle is defined by an effective diffusion coefficient
\begin{equation}
\begin{split}
\hat{D}_\text{eff} = (1-f) \hat{D} + f ,
\label{eqn:deff}
\end{split}
\end{equation}
which in the limit of $\hat{t}^*\ll 1$ is dominated by the term corresponding to bidirectional active walks ($\hat{D}_\text{eff} \rightarrow f$).

The relative importance of processive versus diffusive transport over a length-scale $x$ can be characterized by the P\'eclet number~\cite{godec2015signal}: $\text{Pe}(x) = v x/ D$, which is a dimensionless quantity often used to compare the contributions from advection and diffusion for particles in a flowing fluid~\cite{leal2007advanced}. A large P\'eclet number $\text{Pe}\gg 1$ corresponds to transport that is dominated by the processive motion. In the case where active motion remains processive only up to distances comparable to the run length $\ell$, the relevant P\'eclet number for long-range transport is  $Pe(\ell) = 1/\hat{D}$. Our dimensionless diffusion constant thus describes the relative contribution of diffusion above processive motion over a length scale comparable to the average run-length. 
For the remainder of the discussion, we focus on the regime where the transition time $\hat{t}^*= \frac{2\hat{D}}{\hat{\gamma}} < 1$, so that a distinct regime of processive motion appears between the regimes dominated by passive diffusion and effectively diffusive bidirectional walks. This is the case for the organelle transport examples listed in Table~\ref{tab:paramlist}. We note in passing that the presence of a discernible processive motion is key to identifying active runs in experimental particle-tracking data\cite{chen2013diagnosing,lin2016active,ahmed2014active}, so that systems not in this regime are unlikely to be selected for studies of active transport.


\subsection{Particle spreading: range}
\label{subsec:range}

An alternate metric for the efficiency of particle spreading is the overall range -- or the average size of the domain that has been explored by a halting creeper particle after an interval of time. For a one-dimensional model, the range of each particle is given by its maximum position minus its minimum position over the course of its trajectory. As will be discussed further in Section~\ref{sec:targetsearch}, the range is directly related to rate at which a dispersed population of particles first encounters a target.

Our model permits calculation of the range over time for a halting creeper using the renewal equation method~\cite{feller2015integral,campos2015optimal}. Namely, we define the distribution of first passage times to a target at position $x>0$ (for a particle starting in a diffusive state at the origin) as $F_D(t;x) = F_{D+}(t; x) + F_{DD}(t;x)$, where $F_{D+}$ gives the distribution of first passage times for the fraction of particles that arrive at the target while walking in the positive direction and $F_{DD}$ gives the distribution of times for particles arriving in the passive diffusive state. Similarly, we consider the components of the propagator function defined in Appendix~\ref{app:propagator}, where $G_{DW}(x;t)$ gives the spatial distribution at time $t$ of particles that began in a diffusive state at the origin at time $0$, and are found in the actively walking state at time $t$. The other components $G_{DD}, G_{WD}, G_{WW}$ are defined analogously, with additional expressions for $G_{+D}, G_{+W}$ giving the propagator for particles that are initially walking in the positive direction, and end up in either the diffusive or the actively walking state. One of the renewal equations for this system is then given by,
\begin{equation*}
\begin{split}
G_{DD}(x;t) =  \int_0^t dt' & \left[ 
F_{DD}(t'; x)G_{DD}(0^+; t-t') \right. \\
+ & \left. F_{D+}(t'; x)G_{+D}(0^+; t-t')  \right],
\end{split}
\end{equation*}
where $G_{+D}(0^+; t') = \lim\limits_{~~\epsilon\rightarrow 0^+} G_{+D}(\epsilon; t')$.
This expression describes a convolution between the probability that the particle first hits the target $x$ at time $t'$ and then returns to position $x$ within the remaining time $t-t'$. Analogous renewal equations are derived for $G_{DW}, G_{WD}, G_{WW}$. After a Laplace transform in time, this convolution structure can be expressed as a product, which yields a system of equations,
\begin{equation}
\begin{split}
\left[
\begin{array}{cc}
\widehat{G}_{DD}(0^+) & \widehat{G}_{+D}(0^+) \\
\widehat{G}_{DW}(0^+) & \widehat{G}_{+W}(0^+)
\end{array}
\right] 
\left[\begin{array}{c}
\widehat{F}_{DD} \\
\widehat{F}_{D+}
\end{array}\right]
 = \left[\begin{array}{c}
 \widehat{G}_{DD}(x) \\
 \widehat{G}_{DW}(x) 
 \end{array}\right].
\end{split}
\label{eqn:renewD}
\end{equation}
This system can be solved to calculate the first passage time $\widehat{F}_D$ for particles that started in the diffusive state. An analogous system yields the first passage time for particles that began in the active state:
\begin{equation}
\begin{split}
\left[
\begin{array}{cc}
\widehat{G}_{DD}(0^+) & \widehat{G}_{+D}(0^+) \\
\widehat{G}_{DW}(0^+) & \widehat{G}_{+W}(0^+)
\end{array}
\right] 
\left[\begin{array}{c}
\widehat{F}_{WD} \\
\widehat{F}_{W+}
\end{array}\right]
= \left[\begin{array}{c}
\widehat{G}_{WD}(x) \\
\widehat{G}_{WW}(x) 
\end{array}\right].
\end{split}
\label{eqn:renewW}
\end{equation}

The range of the halting creeper particles over time ($Z(t)$) can be related to the Laplace-transformed first passage time $\widehat{F}(s; x) $ according to~\cite{campos2015optimal},
\begin{equation}
\begin{split}
Z(t) = \mathcal{L}^{-1} \left[\frac{1}{s} \int_{-\infty}^\infty \widehat{F}(s; x) dx  \right],
\end{split}
\label{eqn:rangeEq}
\end{equation}
where $\widehat{F}$ is a linear combination of $\widehat{F}_D$ and $\widehat{F}_W$, weighted by the equilibrium probability that the particle starts in an active or a passive state.
To calculate the range function, we analytically perform the Fourier inversion of the propagators $\widehat{G}_{DD}(0^+), \widehat{G}_{+D}(0^+),\widehat{G}_{DW}(0^+), \widehat{G}_{+W}(0^+) $. The spatial integral over $x$ results in the right hand side of the renewal equation being expressed as $\widehat{\widetilde{G}}_{DD}(k=0)$, etc. 
While short-time and long-time limits of the range can be obtained directly from the large $s$ and small $s$ limits of the renewal equations, the relevant time-scales for biological processes can span across many orders of magnitude, thus making it desireable to calculate the particle spreading efficiency over all time scales.
To this end, we invert the Laplace transform numerically using Talbot's algorithm~\cite{talbot1979accurate}.

\begin{figure*}[t!]
	\includegraphics[width=\textwidth]{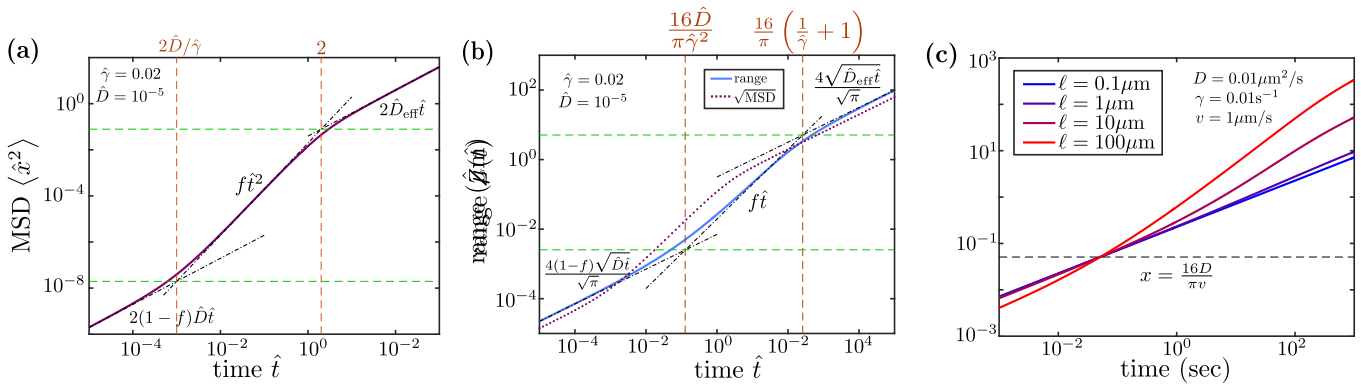}
	\caption{Contribution of passive and active motion to spreading of particles at different length and time scales. (a) Mean squared displacement for halting creepers. Black dash-dotted lines show scaling regimes. Vertical dashed lines indicate transition times between the regimes. Horizontal dashed lines indicate transition length scales. (b) Dimensionless range versus time for a halting creeper particle, with scaling regimes, transition times, and transition length scales indicated. Dotted curve shows the root mean squared displacement for comparison. (c) Range versus time for typical parameter values for intracellular organelles, showing the increase in long-range transport with increasing run length, above a length scale indicated by the black dashed line. }
	\label{fig:range}
\end{figure*}

 As shown in Fig.~\ref{fig:range}, the range exhibits similar transitions in scaling as the MSD. However the transitions between the different regimes are shifted to longer times. In the short time limit, the average range of particles with an equilibrium initial distribution between active and passive states is given (in dimensionless units) by
\begin{equation}
\begin{split}
Z(\hat{t}) \rightarrow 4(1-f)\sqrt{\frac{\hat{D}}{\pi}} \sqrt{\hat{t}} + f\hat{t}.
\end{split}
\label{eqn:zapprox}
\end{equation}
Thus, the ballistic motion dominates over the diffusive motion above a critical transition time
\begin{equation}
\begin{split}
\hat{t}^*_\text{range} = \frac{16 \hat{D}}{\pi \hat{\gamma}^2}.
\end{split}
\end{equation}
In the case where particles spend very little time in active motion ($\hat{\gamma} \ll 1$), this time-scale is substantially longer than the transition time $\hat{t}^*$ for the MSD. The corresponding length scale for the transition from primarily diffusive to primarily ballistic motion is
\begin{equation}
\begin{split}
\hat{x}^*_\text{range} = \frac{16\hat{D}}{\pi\hat{\gamma}(1+\hat{\gamma})}.
\end{split}
\label{eqn:xstar}
\end{equation}

At longer times, there is a subsequent transition from the ballistic scaling of the range to the effectively diffusive long-time scaling,
\begin{equation}
\begin{split}
Z(\hat{t}) & \rightarrow 4\sqrt{\frac{\hat{D}_\text{eff}}{\pi}} \sqrt{\hat{t}}, \quad \hat{t}\gg \hat{t}^{**}_\text{range},
\end{split}
\end{equation}
which occurs at a secondary transition time $t^{**}_\text{range}$ and corresponding length scale $x^{**}_\text{range}$ given by,
\begin{equation}
\begin{split}
\hat{t}^{**}_\text{range} & = \frac{16\hat{D}_\text{eff}}{\pi f^2}, \\
\hat{x}^{**}_\text{range} & = \frac{16}{\pi} \(1 + \frac{\hat{D}}{\hat{\gamma}}\).
\end{split}
\end{equation}

In the case of small fraction of time spent walking, this transition time is again shifted substantially above what would be expected from the MSD behavior, where the corresponding transition occurs at $\hat{t}^{**} =2$. In the limit $\hat{D}/\hat{\gamma}\ll 1$, the transition time for the range can also be expressed as $\hat{t}^{**}_\text{range}=\frac{16}{\pi}(1+1/\hat{\gamma})$, comparable to the cycle time required for a single particle to transition between an active and a passive state and back again.

This result highlights the fundamental insufficiency of the MSD in describing the efficiency with which the particles explore their domain. Specifically, for a very small equilibrium walking fraction $f$, the time required for the active walks to contribute substantially to the average range can be well above the time-scale $1/\gamma$ for an individual particle to start walking. Similarly, in this regime the range will only exhibit diffusive scaling at time-scales long enough for individual particles to execute multiple starting and stopping transitions.  Examples of particle motion where the pause time  substantially exceeds the processive run time include organelles (such as peroxisomes) whose active transport is mediated by hitch-hiking on other organelles~\cite{salogiannis2016peroxisomes}, and particles whose motion is driven by hydrodynamic entrainment due to cytoplasmic flow associated with nearby passing particles~\cite{mussel2014drag}. In such cases, the MSD does not accurately represent the rate at which these particles explore their domain.

We note that in the case where $\hat{D}<1$, which corresponds to most biologically relevant examples, increasing the run length ({\em e.g.:} by decreasing the stopping rate $\lambda$) raises the particle range for all length scales above $x > \frac{16 D}{\pi v}$ (Fig.~\ref{fig:range}), corresponding to the length at which the Peclet number $\text{Pe}(x)$ becomes substantial.  The implication is that longer processive runs improve the ability of particles to explore their domain at all length scales where active walks move faster than diffusion.

\section{Particle dispersion through bidirectional transport}
\label{sec:dispersion}

\begin{figure*}[t!]
	\centering
	\includegraphics[width=\textwidth]{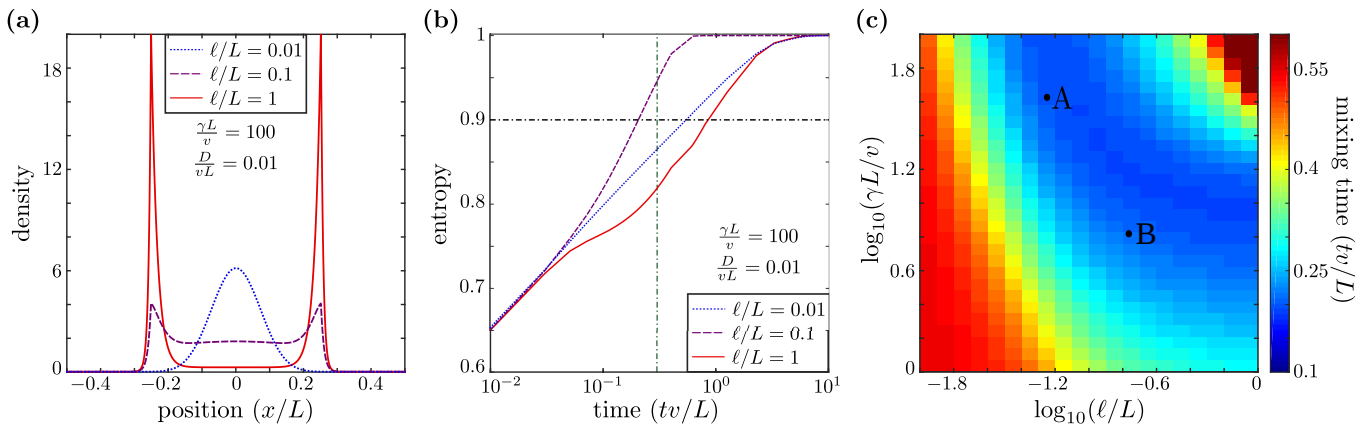}
	\caption{\label{fig:dispersion} Dispersion of particles towards a uniform distribution via bidirectional transport. All length units are non-dimensionalized by domain length $L$ and all time units by $L/v$. (a) Particle distribution density for different run lengths, at dimensionless time $0.3$.
		  (b) Entropy vs time for different run lengths. 
		  The horizontal dash-dot line denotes the threshold entropy for the system to be considered well-mixed. The vertical green dash-dot line is at dimensionless time 0.3 (c) Time to reach a well-mixed state as a function of the run length ($\ell$) and rate of transition to an active state ($\gamma$). Points A and B are drawn at corresponding transport parameters for lysosomes in monkey kidney cells, and PrP$^\text{C}$ vesicles in mouse axons, respectively (Table~\ref{tab:paramlist}). }
\end{figure*}

Having established the speed of particle spreading via multi-modal bidirectional transport, we now turn to consider explicitly the efficiency with which such transport can achieve a particular cellular goal. Certain metabolic and regulatory needs of the cell require a well-dispersed distribution of organelles throughout the cell interior. For instance, mitochondria are found throughout neuronal axons, providing a locally available energy source through glucose metabolism~\cite{pekkurnaz2014glucose}. In fungal hyphae, peroxisome organelles are maintained at near uniform distribution\cite{lin2016active}, allowing for rapid neutralization of toxic metabolic byproducts~\cite{singh1997biochemistry}.

Establishing a well-mixed distribution relies not only on the ability of particles to move rapidly through the cell, but also on the ability of a transport mechanism to disperse and flatten regions of highly concentrated particles. We focus specifically on the rate with which a bolus of particles is spread over a cellular region. Such a process becomes necessary, for instance, in the case of rapid organelle production in response to an external signal, where the organelles must then be spread through long cellular projections such as axons or hyphae. 

We use the halting creeper model to explore how different transport parameters affect the efficiency of such dispersion.
%
%
Because we are interested in the initial establishment of an equilibrium spatial distribution, we consider particles that originate at $x=0$ in the passive state, whose distribution is given by $G_D(x,t) = G_{DD}(x,t)+G_{DW}(x,t)$. This function can be evaluated by numerical Fourier-Laplace inversion of the transformed distribution, as described in Appendix~\ref{app:propagator}. The time evolution of the distribution is plotted in Fig.~\ref{fig:dispersion}a.
  Note that 
 long walk lengths result in little dispersion of particles, with the distribution splitting into two narrow, processively moving peaks. Short walk lengths lead to an effectively diffusive motion, with the particle distribution assuming the form of a slowly spreading Gaussian. An intermediate walk length combines both the rapid spreading of the distribution  with the flattening of localized peaks to enable more efficient dispersion. The limits for large and small walk lengths suggest that there exists an optimal run length $\ell$ for which the particles are most efficiently mixed.

 \begin{figure*}[t!]
 		\includegraphics[width=0.9\textwidth]{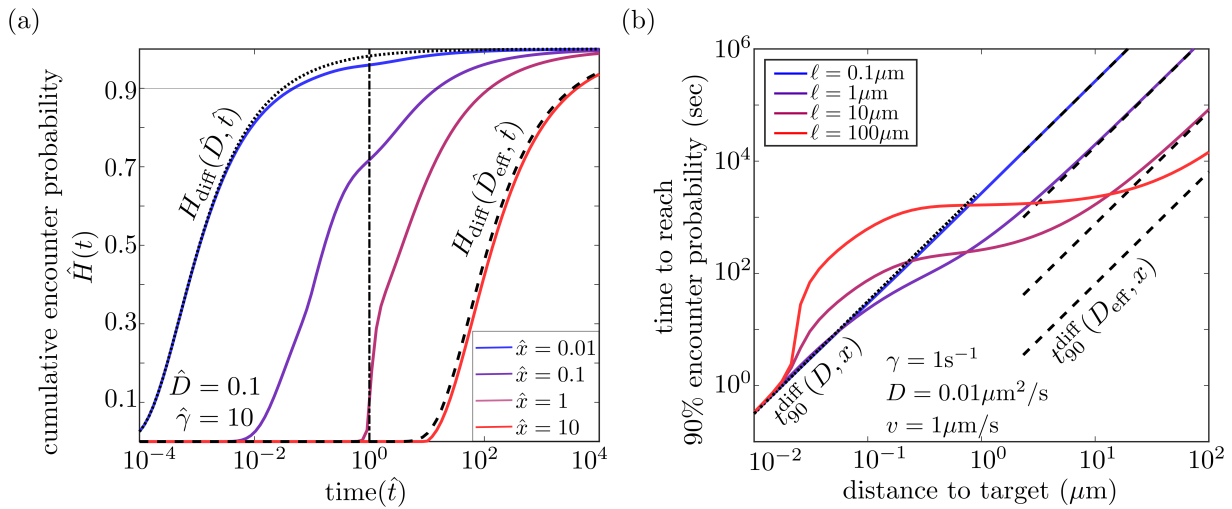}
 		\caption{Target capture times for a single particle. (a) Cumulative encounter probability for different initial distances to the target ($\hat{x}$). The dotted and dashed lines denote the encounter probability to a target at distance $\hat{x}=0.01$ by diffusive particles with diffusivity $\hat{D}$, and to a target at distance $\hat{x}=10$ by diffusive particles with diffusivity $\hat{D}_\text{eff}$, respectively.
 			 The dash-dot line denotes the average time required for a particle in the active state to cover a distance of $\hat{x} = 1$. All length units are non-dimensionalized by $\ell$ and all time units by $\ell/v$.  (b) Time to reach $90\%$ capture probability for different run lengths, assuming rapid starting rate $\gamma=1\text{s}^{-1}$ and distances appropriate for intracellular organelle transport. The dotted and dashed lines denote $t_{90\%}$ for diffusive particles with diffusion coefficient $D$ and $D_{\mathrm{eff}}$, respectively. }
 		\label{fig:singleFPT}
 \end{figure*}
 
A number of different metrics have been developed for quantifying the rate of mixing driven by stochastic processes~\cite{danckwerts1952definition,stone2005imaging,thiffeault2012using}, including several that track the approach of a bolus of particles towards uniform spread~\cite{ashwin2002acceleration,camesasca2006quantifying}. A commonality of these measures is their dependence on a particular length scale of interest~\cite{thiffeault2012using} over which particles are to be mixed. For our one-dimensional system, we introduce a length $L$ corresponding to the size of the domain on which uniform distribution is desired. This length represents the approximate extent of the tubular cell region across which particles are being dispersed. It can range over  many orders of magnitude, with mammalian axons reaching up to a meter in length. Example values for some cellular systems are listed in Table~\ref{tab:paramlist}.
We calculate the spatial distribution of halting creeper particles originating in the center of a domain of length $L$ with reflecting boundary conditions, implemented using the standard image method~\cite{redner2001guide}. The mixing of the particles is quantified via the Shannon entropy of the distribution~\cite{ogawa1975definition,camesasca2006quantifying}, defined as

\begin{equation}
S = -\sum_{i=1}^N \frac{p_i \log(p_i)}{\log(N)},
\label{eq:entropy}
\end{equation}
where the domain is broken up into $N$ bins, and $p_i$ is the probability of a particle being located in bin $i$. Optimal mixing is achieved when the organelles are uniformly distributed, in which case $p_i = \frac{1}{N}$ and $S=1$. Conversely, a distribution with all particles in a single section is the least mixed state, with $S=0$. The entropy has an inherent dependence on the number of bins used for discretizing the probability distribution, and we employ $N=5000$ throughout our calculations.


The time evolution of the entropy is dependent on the dimensionless run-length $\ell/L$ (Fig.~\ref{fig:dispersion}b), with long runs corresponding to an initially slow rise in entropy as the bolus of particles evolves into two coherent spatial peaks until sufficient reversals are achieved to disperse the particles throughout the domain. Short run lengths limit the rate of entropy increase over long times, because the particle distribution spreads slowly as an effectively Gaussian peak. We consider the system to be well-mixed when the entropy crosses a threshold value $S_{t} = 0.9$ and define the time taken to reach this state as the mixing time $t_{mix}$. This mixing time depends in a non-monotonic fashion on both the starting rate ($\gamma$) and run length ($\ell$) of processive walks (Fig.~\ref{fig:dispersion}c). High values of $\gamma$, corresponding to particles that spend most of their time in the active state, give rise to an optimum run length to achieve the most rapid mixing. This effect arises from the need to reverse active walking direction in order to efficiently disperse particles within the domain. However, each such reversal necessitates a waiting time of $1/\gamma$ during which the particles are in a passive state and spreading very slowly. Consequently, at low values of $\gamma$ mixing is most efficiently achieved by particles that carry out very long walks.
 The results shown in Fig.~\ref{fig:dispersion} assume a small value of passive diffusivity ($\frac{D}{Lv} = 0.01$). Increasing this diffusivity would lead to a monotonic rise in the entropy, as diffusion enhances the particle mixing.

\section{Target Search by Multimodal Transport}
\label{sec:targetsearch}

\begin{figure*}
	\begin{minipage}{0.65\textwidth}
		\includegraphics[width=\textwidth]{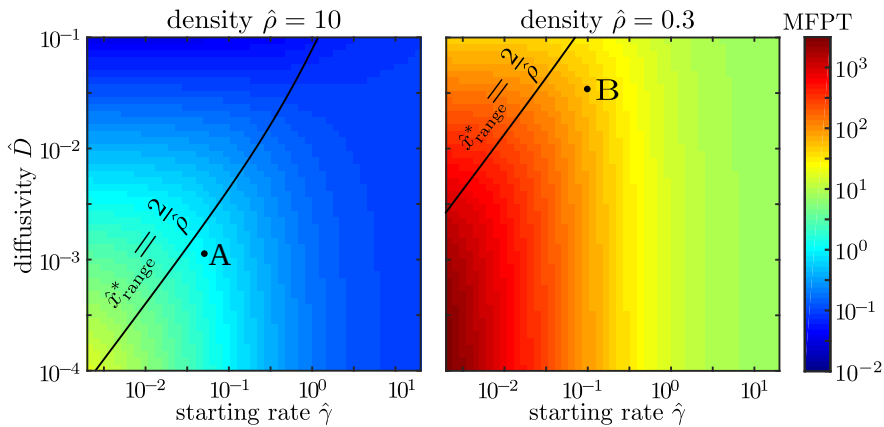}
	\end{minipage}
	~~
	\begin{minipage}{0.30\textwidth}
		\caption{Average time for target capture by  a population of uniformly distributed particles, with two different densities (left: $\hat{\rho} = 10$ and right: $\hat{\rho} = 0.3$). All units are non-dimensionalized by run length $\ell$ and run time $\ell/v$. Marked points show estimated parameters for two cellular systems (A: peroxisome transport in fungal hyphae~\cite{salogiannis2016peroxisomes}, and B: vesicle transport in {\em Aplysia} axons~\cite{ahmed2014active}; see Table~\ref{tab:paramlist}). Black lines mark the transition between diffusion-dominated and transport-dominated motion on the length scale of inter-particle distance (Eq.~\ref{eqn:rhocrit}).}
		\label{fig:capturetimes}
	\end{minipage}	
\end{figure*}

\subsection{Search by a single particle}
In addition to achieving uniform dispersion of particles, another goal of intracellular transport is to deliver organelles to specific cellular regions.  This transport objective arises, for instance, 
 when synaptic vesicles must  reach the presynaptic bouton of a neuron~\cite{hirokawa2005molecular}. 
Using our one-dimensional halting creeper model, we consider the first passage time of a single particle towards a stationary target located at distance $x$.  For simplicity, we consider the case where $x$ is much smaller than the overall extent of the domain, so that the distance to the target $x$ and the processive run length $\ell$ are the only relevant length scales in the problem. As in the case of our dispersion calculations (Sec.~\ref{sec:dispersion}), we consider particles that are initially in the passive state, as applicable to the distribution of newly synthesized organelles.

The distribution of first passage times can be obtained from the renewal equation Eq.~\ref{eqn:renewD}, by carrying out analytic Fourier inversion followed by numerical Laplace inversion of the propagators (see Appendix \ref{app:propagator}).
 The cumulative distribution of encounter times to the target is plotted in Fig.~\ref{fig:singleFPT}a, showing the transition from a passively diffusive process at small distances ($\hat{x} < x^*_\text{range}$) to an effectively diffusive process (with diffusivity $\hat{D}_\text{eff}$) at distances much longer than the run length ($\hat{x} > 1$). Intermediate distances show a sharp increase in the cumulative probability of target encounter at time $\hat{t} = \hat{x}$, corresponding to the arrival of the first processively walking particles.

 Because the average first passage time of a random walk in a semi-infinite domain diverges~\cite{redner2001guide}, we focus on the time required for particles to hit the target with sufficiently high probability. Analogously to our calculations of particle dispersion in Section~\ref{sec:dispersion}, we define the hitting time $t_{90\%}$ as the time by which there is a $90\%$ chance that the particle has hit the target. We note that $t_{90\%}$ is well-defined even on a semi-infinite domain due to the recurrent nature of random walks in one dimension, ensuring a finite hitting time for all particles~\cite{redner2001guide}. For short distances, the time for probable encounter of the particle scales as expected for purely diffusive motion with diffusivity $\hat{D}$ (Fig.~\ref{fig:singleFPT}b),
 \begin{equation}
 \begin{split}
 t_{90}^{\text{diff}}(\hat{x}; \hat{D}) = \frac{\hat{x}^2}{4\hat{D} \left[\text{erf}^{-1}(0.1)\right]^2}
 \end{split}
 \end{equation}
 where $\text{erf}^{-1}$ is the inverse of the error function. At long times, a similar scaling is observed with effective diffusivity $\hat{D}_\text{eff}$.

As is the case when the transport objective is to achieve a uniform distribution of particles, increasing the length of processive runs does not necessarily result in more efficient transport. This is true despite the fact that, unlike previous models of multimodal transport~\cite{loverdo2008enhanced,benichou2011intermittent,godec2015signal}, we consider our particles capable of accessing their target in both the passive and active states. A run length that is much longer than the distance to the target can hinder particle delivery, because particles have a 50\% chance of initiating their motion in the wrong direction. They then require a long time to stop, turn, and return towards the target. At the same time, very short processive runs decrease the overall rate of spread for the particle distribution and thus slow down the target encounter.
 These two effects give rise to an optimum in the efficiency of target delivery, with minimal values of target hit time $t_{90\%}$ occuring at intermediate run lengths $\ell$ (Fig.~\ref{fig:singleFPT}b). This effect is a direct analogue to the optimum walk-length for achieving uniform distribution. The existence of this optimum walk length has also previously been noted for creeper models without any paused or passive state~\cite{campos2015optimal}. 

\subsection{Search by a population of particles}
A closely related objective of intracellular transport is the capture of a target by any one of many moving particles. In this case we assume particles that are initially uniformly distributed with some density $\rho$, and consider the mean first passage time (MFPT) for the first of them to hit the target. Some biological examples include the clearance of toxic cytoplasmic metabolites by any one of a uniformly scattered field of peroxisome organelles~\cite{singh1997biochemistry}, the influx of peroxisomes to plug septal holes in damaged fungal hyphae~\cite{jedd2000new}, or the arrival of lysosomes to fuse with a phagosome and digest its engulfed contents~\cite{bright2005endocytic}. For simplicity, we assume the target is itself immobile and must wait for the particles to come to it via some combination of active and passive transport. In this situation, the relevant length scale is defined by the typical initial spacing between the particles ($1/\rho$). In the limit of a uniform distribution over a very long domain, the MFPT can be related directly to the range of the moving particles~\cite{campos2015optimal}. Specifically, the mean first passage time is given by
\begin{equation}
\begin{split}
\text{MFPT} = \int_0^\infty e^{-\rho Z(t)} dt
\end{split}
\label{eqn:mfpt}
\end{equation}
where $Z(t)$ is the average range of particles over time $t$.
  The behavior of the MFPT is dictated by the dimensionless length scale for the separation between particles [$1/\hat{\rho} = 1/(\rho \ell)$]. When this length scale is short enough that active walks remain processive ($1/\hat{\rho} \ll \hat{x}^{**}_\text{range}$), we can approximate the particle range as a linear combination of a diffusive and a ballistic process (Eq.~\ref{eqn:zapprox}). The integral for the MFPT can then be approximated analytically as
\begin{equation}
\begin{split}
\text{MFPT} \approx \frac{1}{f\hat{\rho}} - \sqrt{\frac{4\hat{D}}{\hat{\gamma}^2\hat{\rho}f}}\exp\left[\frac{\hat{x}^*_\text{range}\hat{\rho}}{4}\right] \text{erfc}\left[\frac{\hat{x}^*_\text{range}\hat{\rho}}{4}\right],
\end{split}
\end{equation}
where erfc is the complementary error function. The limits for high and low particle concentration are given by
\begin{equation}
\begin{split}
\text{MFPT} & \rightarrow \frac{\pi(1+\hat{\gamma})^2}{8\hat{D}\hat{\rho}^2}, \quad \frac{1}{\hat{x}^*_\text{range}}\ll \hat{\rho}\ll\frac{1}{\hat{x}^{**}_\text{range}},\\
\text{MFPT} & \rightarrow \frac{1}{f\hat{\rho}}, \quad  \hat{\rho}\ll\frac{1}{\hat{x}^{*}_\text{range}},
\end{split}
\end{equation}
where the high density limit corresponds to diffusive scaling of MFPT with the distance between particles while the low density limit corresponds to ballistic scaling. 
Setting these two limits equal to each other indicates that a transition in the encounter times occurs at a critical length scale
\begin{equation}
\begin{split}
\frac{1}{\hat{\rho}_\text{crit}} = \frac{\hat{x}^*_\text{range}}{2},
\end{split}
\label{eqn:rhocrit}
\end{equation}
 which can be equivalently expressed as
\begin{equation}
\begin{split}
f\text{Pe}\(\frac{1}{\hat{\rho}}\) = \frac{8}{\pi(1+\hat{\gamma})^2}.
\end{split}
\label{eqn:rhocrit2}
\end{equation} 
This transition corresponds to a particle density where processive walks begin to dominate the ability to rapidly encounter targets, which occurs when the P\'eclet number for the distance between particles, multiplied by the fraction of time spent walking, is of order unity.

 A calculation of the mean first passage time accurate at all length scales can be carried out by numerical inversion of the Laplace-transformed range function (Sec.~\ref{subsec:range}), and the results are plotted in Fig.~\ref{fig:capturetimes} for two values of particle density. The black line indicates the transition between behavior dominated by diffusive versus by processive particle motion (Eq.~\ref{eqn:rhocrit}). Below this line, active transport dominates the motion of the particles and the time to reach the target is insensitive to the passive diffusivity. Above this line, passive diffusion dominates and the target search is insensitive to the fraction of time that the particles spend in processive motion. The parameters relevant to two example biological systems (peroxisome transport in fungal hyphae and vesicle transport in {\em Aplysia} neurons) are marked with dots. 
 

The two example cases fall  near the transition region, where both passive diffusion and active processive walks contribute to the ability of these organelles to reach any target position within the cell. While previous modeling studies have indicated that both transport mechanisms are important to the maintenance of a uniform distribution of peroxisomes in hyphae~\cite{lin2016active}, we demonstrate here that the particle density falls in an intermediate regime such that diffusion and 
 active walks both contribute to efficient target search by the population of peroxisomes.


\section{Transport in a Tube and the Benefits of Tethering}
\label{sec:tethering}
\begin{figure*}[ht!]
	\centering
	\includegraphics[width=\textwidth]{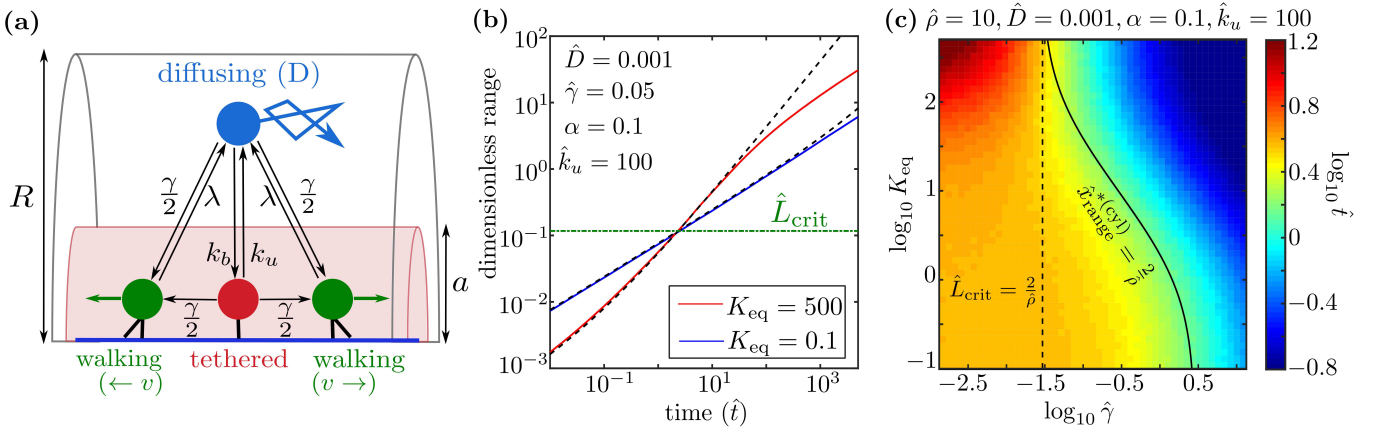}
	\caption{\label{fig:tethering} Effects of tethering on transport. a) Schematic for the model of a halting creeper in a cylinder, with tethering. The smaller cylinder denotes the region within which particles can tether to the microtubule or initiate active transport. Rates of transition between states are labeled with corresponding arrows. b) Range vs time for weak ($K_\text{eq}=0.1$) and strong ($K_\text{eq}=500$) tethering. Dashed black lines show analytical approximations in the limits of no tethering and infinitely strong tethering, accurate for short to intermediate times (Eq.~\ref{eqn:zapproxcyl}). Horizontal dash-dotted line indicates the transition length-scale $\hat{L}_\text{crit}$ where tethering becomes advantageous.  c) Average time for target capture by a population as a function of the starting rate $\gamhat$ and binding strength $K_\text{eq}$.  The solid line indicates the transition from diffusive to active transport as the dominant transport mode at different values of tethering strength. The dashed line shows the transition where strong tethering becomes advantageous for target encounter.}
\end{figure*}
Active transport in a cell occurs via motor proteins attached to microtubule tracks.
Even very narrow cellular projections are typically substantially wider than the diameter of a single microtubule. Consequently, organelles must navigate transversely through the cytoplasmic environment in order to encounter a microtubule and engage in active processive motion.
 A mechanism to keep organelles located close to the microtubule can improve transport efficiency by reducing this search time. In many cases, organelles are believed to be tethered to the microtubule tracks, preventing them from dissociating and diffusing even when they pause after a processive walk. This tethering can be accomplished by additional inactive motors attached to the organelle~\cite{cooper2009diffusive,ross2008kinesin} or by specific molecular adaptors linking the organelle directly to the microtubule~\cite{kang2008docking,chada2004nerve}. 
 
 It has been speculated that tethering can enhance transport by forcing the organelle to remain in proximity to the microtubule tracks, thereby effectively increasing the rate at which processive walks are initiated~\cite{hancock2014bidirectional}. At the same time, tethering can severely limit the intracellular space that can be explored by an organelle in the passive state, either by reducing the axial diffusivity in the case where inactive motors slide diffusively along microtubules~\cite{cooper2009diffusive,culver2006microtubule}, or by halting it entirely in the case of organelle docking~\cite{kang2008docking}. The benefits of tethering thus depend on the relative balance between active and passive transport, as well as the radial size of the domain around the microtubule, which determines the delay associated with encountering the track. The former aspect is dependent on the length scale over which transport must be achieved, as discussed in the previous sections.

We extend our halting creeper model to a three-dimensional cylindrical domain of radius $R$, wherein active runs can be initiated only within a radius of size $a<R$, corresponding to a small region surrounding a central track. While cellular projections such as hyphae and axons generally have multiple microtubule bundles~\cite{peter2012computational,lin2016active}, this model serves as an approximation where the size of the cylindrical domain sets the cross-sectional density of the microtubule bundles. In addition to bidirectional walking and passive diffusion states, the particles in this extended model can also enter a  tethered state with rate $k_b$ while within the encounter radius $a$. For simplicity, we assume particles in the tethered state are entirely immobilized. The model could be extended in a straight-forward manner to limited but non-zero diffusivity while in the tethered state.
 Exit from the tethered state occurs at rate $k_u$, with the particle unbinding to a uniform radial distribution within the capture radius $a$. A dimensionless binding strength for tethering is defined by $K_\text{eq} = k_b/k_u$.
 
 We note that this model assumes that tethering does not in any way hinder the initiation of an active run, so that particles transition to the active state with the same rate regardless of whether they are bound or freely diffusing within the capture radius. While it is possible for tethering to either speed up or slow down the association of an organelle with a motor or a carrier particle, depending on the length, flexibility, and configuration of the tether, we neglect this effect here. 
Our model for tranport in a cylindrical tube around a microtubule track is summarized schematically in Fig.~\ref{fig:tethering}a. 

In the limit of rapid transverse diffusivity or small domain size ($D/R^2 \gg \gamma, k_b$), diffusive particles remain equilibrated throughout the cross section of the domain, and the effective rates of starting a walk or binding become $\alpha^2 \gamma$ and  $\alpha^2 k_b$, respectively, where $\alpha = a/R$. In this limit, the delays associated with transverse diffusive transport are elimitated, and the equilibrium fraction of particles in each state can be easily calculated. For particles starting at equilibrium, the long-time diffusivity is then given by
\begin{equation}
\begin{split}
\hat{D}_\text{eff} & = \hat{D}f_\text{diff} + f_\text{walk}, \\
f_\text{walk} & = \frac{\hat{\gamma}\alpha^2}
{\hat{\gamma}\alpha^2+\left(\frac{\alpha^2 \keq+\hat{\gamma}/\hat{k}_u+1}{\keq+\hat{\gamma}/\hat{k}_u+1}\right)}, \\
f_\text{diff} & = \frac{\left(\frac{\hat{\gamma}/\hat{k}_u+1}{\keq+\hat{\gamma}/\hat{k}_u+1}\right)}{\hat{\gamma}\alpha^2+\left(\frac{\alpha^2 \keq+\hat{\gamma}/\hat{k}_u+1}{\keq+\hat{\gamma}/\hat{k}_u+1}\right)}
 \\
\end{split}
\label{eqn:defftether}
\end{equation}
where $f_\text{walk}$ and $f_\text{diff}$ are the fraction of particles in the active and diffusive state, respectively. We again non-dimensionalize all length units by the run-length $\ell$ and all time units by the run-time $\ell/v$, for consistency with previous calculations.

In the more general case where the delay due to transverse diffusion is included, it can be shown (see Appendix~\ref{app:derivetether}) that for a particle which begins uniformly distributed in the diffusive state within radius $a$, the mean waiting time to enter a walking state is identical to the fast-diffusion limit and is given by,
\begin{equation}
\begin{split}
\left<\hat{t}_w\right> = \frac{1}{\hat{\gamma}}\left[\frac{\alpha^2 K_\text{eq} + 1 + \hat{\gamma}/\hat{k}_u}{\alpha^2 (K_\text{eq} + 1 + \hat{\gamma}/\hat{k}_u)}\right].
\end{split}
\label{eqn:tstartw}
\end{equation}
This average time ranges from $1/(\alpha^2 \gamma)$ in the limit of low binding strength to $1/\gamma$ in the limit of strong binding, and is independent of the diffusivity $D$. In the case of very slow diffusion, those particles that escape the binding radius $a$ take a long time to return, but such escape before initiating a walk becomes concomitantly less likely, with these two effects canceling each other out in the calculation of the average time to start walking. 
Because particles are assumed to distribute uniformly across radius $a$ when leaving the tethered state, this equivalence of the average time to initiate a subsequent walk means that the long-time behavior of particles matches the fast-diffusivity limit, regardless of the actual value of $D$.

By contrast, we note that the standard deviation in the time required to start a walk, for a particle that starts diffusive and uniformly distributed within $a$, is dependent on the diffusivity (see Appendix \ref{app:derivetether}).
Slow diffusion and strong binding can greatly increase the variance in the time required for a particle to start a walk, leading to large variability in the amount of time individual particles remain in a passive or tethered state over a particular time interval of observation. This extreme case may contribute to the identification of apparently immobile populations of particles observed in some {\em in vivo} organelle tracking studies~\cite{chang2006mitochondrial}.

The effectiveness of tethering in improving transport over a long time can be inferred from the derivative of the effective diffusivity $\hat{D}_\text{eff}$ with respect to the binding strength $K_\text{eq}$. A positive derivative signifies that long-range transport is accelerated by tethering, whereas a negative value indicates that tethering hinders transport. Tethering is advantageous in the long-time limit when the following criterion is satisfied:
\begin{equation}
\begin{split}
(1-\alpha^2)\left(\frac{\hat{\gamma}}{\hat{\gamma}+1}\right) \text{Pe}(\ell)>1.
\end{split}
\label{eqn:tetherlong}
\end{equation}
%
This expression summarizes the idea that tethering is helpful for long-range transport in situations where the domain is wide ($\alpha \ll 1$), where the rate of walking is substantial compared to the pausing rate ($\hat{\gamma}\gg 1$), and where active runs move the particles faster than diffusion over the longest processive length scale ($\text{Pe}(\ell)\gg 1$).

Below the long-time diffusive limit, the extent to which tethering aids transport depends on the length scale of interest. In particular, at times much shorter than the cycle time to initiate and stop an active walk, the dimensionless particle range can be approximated by
\begin{equation}
\begin{split}
Z(\hat{t}) \approx 4 f_\text{diff} \sqrt{\frac{\hat{D}\hat{t}}{\pi}} + f_\text{walk}\hat{t},
\end{split}
\label{eqn:zapproxcyl}
\end{equation}
in a manner analogous to Eq.~\ref{eqn:zapprox}. This expression can be inverted to calculate the time at which a particular range is reached. Comparing the low $K_\text{eq}$ and high $K_\text{eq}$ limits indicates that the ability of particles to tether to the track allows them to explore more rapidly over length scales above
\begin{equation}
\begin{split}
\hat{L}_\text{crit} = x^*_\text{range}\frac{(1+\hat{\gamma})^2}{(1-\alpha^2)^2}.
\label{eqn:Lcrit}
\end{split}
\end{equation}
For large domains ($\alpha\ll 1$) and low propensity for active walking ($\hat{\gamma}\ll 1$), the tethering is helpful over all length scales where processive active motion is the dominant form of transport, as defined by the critical length $\hat{x}^*_\text{range}$ (Eq.~\ref{eqn:xstar}).


We use kinetic Monte Carlo methods to simulate the spreading of particles within our cylindrical model. The simulations are accelerated with the use of analytically calculated Green's functions to propagate the particles within homogeneous cylindrical domains (see Appendix~\ref{app:simdetails}), allowing for efficient sampling of particle behavior over a broad set of parameters. 

The average axial range for a population of particles can be obtained as a function of time from the simulations. Fig.~\ref{fig:tethering}b shows the time evolution of the range for weak and strong tethering. The transport parameters used are relevant for peroxisomes in fungal hyphae (Table~\ref{tab:paramlist}), with the domain width assumed to be $R=1\mu$m and a central region of width $a=0.1\mu$m. For consistency with previous calculations, results are reported in dimensionless units, using the run length ($\ell\approx 7\mu$m) and processive walking time ($1/\lambda \approx 3$~s) as the length and time units. The critical length scale for this system is $\hat{L}_\text{crit} \approx 0.12$, below which the average range for strongly tethered particles is lower than the weakly tethered ones. For length scales above $\hat{L}_\text{crit}$, strongly tethered particles explore over a greater range. The full extent of a hyphal growth tip ($\hat{L}\approx 8$) is several times longer than this critical length scale, highlighting the potential benefit of tethering for distributing peroxisome particles over the entire growth tip.

Having established the length scales over which tethering is advantageous, we now calculate explicitly the effect of tethering on the average search time by a population of particles with dimensionless density $\hat{\rho}$. The capture time is defined as the first passage time to an arbitrary cross-section of the cylinder, by a population of particles equilibrated between states and uniformly distributed along the axis of the cylinder.
A surface plot of the average capture times versus binding strength $K_\text{eq}$  and walking rate $\hat{\gamma}$ is shown in Fig.~\ref{fig:tethering}c. The effect of tethering on the average time to target capture varies depending on $\hat{\gamma}$. 
For particles with a very small probability of engaging in active runs, tethering hinders target search by limiting mobility in the passive state. For particles with a high propensity for active motion, tethering can aid their ability to encounter targets by increasing the amount of time spent in the region where active runs can be initiated. We approximate the parameter regime where this transition occurs by analytically calculating the integral for the MFPT (Eq.~\ref{eqn:mfpt}), using the short time approximation of the particle range (Eq.~\ref{eqn:zapproxcyl}). Comparing the low $K_\text{eq}$ and high $K_\text{eq}$ limits yields a transition at a critical particle density
\begin{equation}
\begin{split}
\hat{\rho}_\text{tether} = \(\frac{2}{\hat{L}_\text{crit}}\)\(\frac{1+\alpha^2\hat{\gamma}}{1-\alpha^2}\)^2.
\end{split}
\end{equation}
For small values of $\alpha, \hat{\gamma}$, this transition is equivalent in form to the critical length scale where processive walks first begin to play an important role, as calculated in Eq.~\ref{eqn:rhocrit}. For parameters relevant to the motion of peroxisomes in fungal hyphae, we compare the critical particle density ($\hat{\rho}_\text{tether}\approx 17$) with the observed density of peroxisomes ($\hat{\rho}\approx 10$). Because the observed density is comparable to the critical density, we expect that tethering would have only a weak benefit on the ability of the peroxisomes to patrol the cytoplasm and encounter targets within the cell. 

For a given finite binding strength $K_\text{eq}$, the MFPT to the target will be dominated by either diffusive or processive motion, depending on the fraction of particles in each state. The transition to the regime where encounter times are sensitive to the initiation of active walks occurs when the spacing between particles hits a critical length scale where such walks between to dominate. This length can be obtained analogously to the expression for $\hat{x}^{*}_\text{range}$ (Eq.~\ref{eqn:xstar}) by
replacing the starting rate $\hat{\gamma}$  with an effective starting rate based on the average time to initiate a walk: $\hat{\gamma}_\text{eff} = 1/\left<t_w\right>$. In the case of rapid binding/unbinding ($\hat{k}_u \gg \hat{\gamma}$), this rate is approximated as
\begin{equation}
\begin{split}
\hat{\gamma}_\text{eff}  \approx \hat{\gamma} \left[\frac{\alpha^2(K_\text{eq} +1)}{\alpha^2 K_\text{eq} + 1}\right]
\end{split}
\end{equation}
The critical particle density is then given by
\begin{equation}
\begin{split}
\hat{\rho}^{\text{(cyl)}}_\text{crit} = \frac{2}{\hat{x}^{* (\text{cyl})}_\text{range}} = \frac{\pi \hat{\gamma}_\text{eff}}{8\hat{D}}, 
\end{split}
\end{equation}
where $\hat{x}^{* (\text{cyl})}_\text{range}$ is the length scale for transition between diffusive and processive motion in the model of a halting creeper within a cylindrical domain. This transition is shown with a solid black line in Fig.~\ref{fig:tethering}c.
\section{Summary}
\label{sec:conclusion}

We have employed a simplified ``halting creeper" model, consisting of stochastic interchange between passive diffusion and active processive walks, to investigate the efficiency of transport within an extended cylindrical domain. Specifically, this model is applicable to the transport of organelles within long narrow cellular processes such as neural axons and fungal hyphae. We explore the space of relevant parameters, including the rates of transition between passive and active states and the relative speed of diffusion versus active transport, as characterized by the P\'eclet number over different length scales. Our results highlight the importance of the relevant length scale in determining the contributions of the different transport modes and we identify simple expressions for the time [$t_\text{range}^* = \frac{16D\lambda^2}{\pi v^2 \gamma^2}$] the length [$x_\text{range}^* = \frac{16D\lambda^2 }{\pi v \gamma (\lambda+\gamma)}$] at which processive motion dominates particle spreading. We emphasize the use of the average range as a metric for the ability of particles to explore their domain via multimodal transport, demonstrating passive diffusion can play an important role over longer length scales than expected based on the classic analysis of the mean squared displacement.

We focus specifically on the contributions of active and passive transport to several key objectives relevant to the cell. First, we consider the establishment of a uniform distribution from a bolus of particles, demonstrating that efficient dispersion is achieved at intermediate run-lengths that can be substantially smaller than the domain size. This result indicates the importance of bidirectional active transport with frequent reversals in the movement of particles that must be spread broadly throughout a large domain, as is the case with metabolic organelles such as peroxisomes and mitochondria. Second, we quantify the rate at which a single particle first encounters a stationary target, showing again an advantage to intermediate run lengths that minimize the time wasted pursuing a long processive walk in the wrong direction. Third, we consider the rate of encounter to a target by the first of a population of halting creeper particles, identifying the parameter regime where active transport or diffusion dominate the motion, and showing that several examples of biological interest fall in the intermediate regime where both modes of transport contribute substantially to target encounter.

Finally, we investigate an extension of the one-dimensional model to a cylindrical domain, where active transport can only occur in a narrow region along the axis and where particles can enter a halted tethered state that both enhances the effective rate of initiating an active run and limits their ability to explore the domain while in the passive state. The advantages of tethering to microtubule tracks have been a topic of much speculation in the literature on intracellular transport~\cite{hancock2014bidirectional,cooper2009diffusive,culver2006microtubule}. We delineate the parameter regime in which tethering is expected to aid the long-time dispersion of particles (Eq.~\ref{eqn:tetherlong}) and identify a critical length scale $L_\text{crit}$ (Eq.~\ref{eqn:Lcrit}) below which tethering hinders the ability of the particles to explore their domain. For several example intracellular transport systems (Table~\ref{tab:paramlist}), this critical length is on the order of a few hundred nanometers, confirming the advantages of tethering for transport over many micron length scales. 

The results derived in this work highlight the complementary role of diffusion and processive transport in fulfilling cellular goals for delivering and distributing cytoplasmic organelles. The derived expressions can be employed for analyzing data on measured transport parameters to determine the length scales and transport objectives where active motor-driven motion is expected to dominate, where bidirectional transport with limited processivity is advantageous, and where tethering to cytoskeletal tracks can aid overall organelle dispersion.

\section{Acknowledgements}
\label{sec:acknowledgements}
We thank A. Agrawal and C. Niman for helpful comments on the manuscript and S. Reck-Peterson for fruitful discussions.

\newpage

\appendix 

\section{Propagator for a one-dimensional halting creeper}
\label{app:propagator}

We calculate the position distribution of a particle switching between diffusive transport with diffusivity $D$ and processive motion 
 with speed $v$. Switching between states is a Poisson process with rate $\gamma$ for entering an active state and rate $\lambda$ for leaving an active state (see Fig.~\ref{fig:schematic}). The overall spatial distribution can be obtained by a convolution of propagators for individual states, summed over all possible state transitions.


Starting at an intial position $x = 0$, the spatial distribution of a diffusive particle at a time t is 
\begin{equation}
R_D(x,t) = \frac{1}{\sqrt{4\pi Dt}}e^{-\frac{x^2}{4Dt}}
\label{eqn:diffR}
\end{equation}

We define the joint distribution that the particle first switches to an active state at time $t$ while at position $x$ by,
\begin{equation}
H_D(x,t) = \gamma e^{-\gamma t} R_D(x,t).
\end{equation}


The position distribution of particles starting at $x=0$ in the active state, moving with a velocity $v$ at time $t$ is given as
\begin{equation}
R_{\pm}(x,t) = \delta(x\mp vt).
\label{eqn:walkR}
\end{equation}

The corresponding joint distribution for the time and location of switching from the active to the passive state is
\begin{equation*}
H_\pm(x,t)=\lambda e^{-\lambda t}\delta(x\mp vt)
\end{equation*}
%


We define a ``step" in the particle's trajectory as a switch from the passive to the active state and back to the passive state again. If the particle starts in the passive state at zero time, the position and time distribution at the end of one such step can be expressed as
\begin{equation}
\begin{split}
M(x,t) = \int_{-\infty}^{\infty}dx'\int_{-\infty}^t dt'~H_D(x',t')\\\times\left[\frac{H_+(x-x',t-t')+H_-(x-x',t-t')}{2}\right],
\end{split}
\label{eqn:M}
\end{equation}

where the first term denotes a particle reaching $x'$ at time $t'$ via diffusion and the second term denotes the particle covering a distance $x-x'$ in the remaining time $t-t'$ by walking, integrated over all values of $x'$ and $t'$.

To get the spatial propagator of a halting creeper particle that both starts and ends in a passive state,
we sum over all possible paths between the active and passive states, convolved with the probability that the particle does not leave the passive state in the final time interval (given by $H_D(x,t)/\gamma$). 
 The resulting expression for the propagator can then be expressed as:
\begin{equation}
\begin{split}
G_{DD}(x,t) & = \frac{H_D}{\gamma} + M*_{x,t}\left(\frac{H_D}{\gamma}\right) \\
& + M*_{x,t} M*_{x,t}\left(\frac{H_D}{\gamma}\right)+ ...,
\end{split}
\end{equation}
where $*_{x,t}$ denotes convolution with respect to $x$ and $t$. The first term in the summation corresponds to a particle that never left the passive state, the second term to a particle that performs a single active step before returning to the passive state, the third term includes two active steps, and so forth.

A Fourier transform in space ($x\rightarrow k, M\rightarrow \widetilde{M}$) and a Laplace transform in time ($t\rightarrow s, M\rightarrow \widehat{M}$)
transform the convolutions in $x$ and $t$ to products of functions in $k$ and $s$ respectively. Applying these transforms to the expression for $G_{DD}(x,t)$ yields a geometric series that sums to
\begin{equation}
\begin{split}
\fl{G}_{DD}(k,s) &= \frac{\left(\frac{\fl{H}_D(k,s)}{\gamma}\right)}{1-\widehat{\widetilde{M}}(k,s)}\\
 &= \frac{(s+\lambda)^2+k^2v^2}{(s+\gamma+Dk^2)\left((s+\lambda)^2+k^2v^2\right)-\gamma\lambda(s+\lambda)}.
\end{split}
\end{equation}

The distributions for the other quantities appearing in Eqs.~\ref{eqn:renewD},~\ref{eqn:renewW} of the main text can be derived similarly. The transformed distributions are:

\begin{equation}
\begin{split}
\fl{G}_{DW} &= \frac{\gamma(s+\lambda)}{(s+\gamma+Dk^2)\left((s+\lambda)^2+k^2v^2\right)-\gamma\lambda(s+\lambda)}\\
\fl{G}_{WW} &= \frac{(s+\gamma+Dk^2)(s+\lambda)}{(s+\gamma+Dk^2)\left((s+\lambda)^2+k^2v^2\right)-\gamma\lambda(s+\lambda)}\\
\fl{G}_{WD} &= \frac{\lambda(s+\lambda)}{(s+\gamma+Dk^2)\left((s+\lambda)^2+k^2v^2\right)-\gamma\lambda(s+\lambda)}\\
\fl{G}_{+D} &= \frac{\lambda(s+\lambda+ikv)}{(s+\gamma+Dk^2)\left((s+\lambda)^2+k^2v^2\right)-\gamma\lambda(s+\lambda)}\\
\fl{G}_{+W} &= \frac{(s+\gamma+Dk^2)(s+\lambda+ikv)}{(s+\gamma+Dk^2)\left((s+\lambda)^2+k^2v^2\right)-\gamma\lambda(s+\lambda)}.\\\\
\end{split}
\end{equation}

A linear combination of these distributions, weighted by the equilibrium fraction of particles in each state is used to derive the overall propagator in Eq.~\ref{eqn:propagator}:
\begin{equation}
\begin{split}
\fl{G} = \frac{\gamma}{\gamma+\lambda} \(\fl{G}_{WD} + \fl{G}_{WW}\) +  \frac{\lambda}{\gamma+\lambda} \(\fl{G}_{DD} + \fl{G}_{DW}\). 
\end{split}
\end{equation}


The expressions obtained can be transformed back to real space and real time by a combination of analytical and numerical methods. To calculate the Laplace-transformed expressions in Eq.~\ref{eqn:renewD},~\ref{eqn:renewW}, and \ref{eqn:rangeEq} we invert the Fourier transform analytically as  $\widehat{G}(x,s) = \frac{1}{2\pi}\int_{-\infty}^{\infty}e^{ikx}\fl{G}(k,s)~dk$. The Laplace transform of the range and first passage time distribution can then be inverted numerically using Talbot's algorithm~\cite{talbot1979accurate}. 


An alternate approach for efficiently calculating the propagator in real-space and real-time is to first invert the Laplace transform analytically using the residue theorem, followed by a numerical integral over $k$ to invert the Fourier transform. We use this approach to calculate the  particle distribution at a high spatial resolution in order to find the entropy over time (Fig.~\ref{fig:dispersion}).


\section{Analytical model for multimodal transport in a cylinder, with tethering}
\label{app:derivetether}

In this section we develop the full analytical model for axial transport in a cylinder of radius $R=1$ for particles capable of passive diffusion with diffusivity $D$, of initiating active processive walks with a rate $\gamma$ while within a region of radius $\alpha$ of the central axis, and of entering a stationary tethered state with binding rate $k_b$ while in the same region. The rate constant for unbinding from a tethered state is $k_u$ and for transitioning between an active walk and passive diffusion is $\lambda$ (see Fig.~\ref{fig:tethering}(a) for illustration of the model). 
For ease of the derivation, all length units in this section as well as Appendix~\ref{app:simdetails} are non-dimensionalized by the cylinder radius $R$ and all time units are nondimensionalized by $R/v$ where $v$ is the processive velocity of actively walking particles. We give our final results in fully dimensional units to facilitate comparison with other sections of the manuscript.

Our model is developed in an analogous manner to the approach previously used for modeling facilitated diffusion by DNA-binding proteins that occurs via a combination of 3D diffusion and 1D sliding along a filament~\cite{koslover2011theoretical}. We describe the particle motion by a system of individual states with Markovian transitions between them. The rates of transition between the states are time-varying, depending specifically on the time interval since the particle first entered the state. These states ( Fig.~\ref{fig:tetherstates}) consist of: a tethered state ($h$), a walking state ($w$), a state ($n$) wherein the particle started at radius $\alpha-\epsilon$ and has remained within a radius $\alpha$, a state ($n_u$) where the particle started uniformly distributed within radius $\alpha$ and has remained within that inner region, a state ($f$) where the particle started at radius $\alpha+\epsilon$ and has remained outside the inner region at a radius greater than $\alpha$ and a state $f_u$ where the particle started uniformly distributed in the outer region and has remained in the outer region. When computing statistics for the overall motion of the particle, we take the limit $\epsilon \rightarrow 0$. 
The axial propagation of a particle in states $n, n_u, f, f_u$ is given by the propagator function for diffusive motion $R_D(x,t)$ (Eq.~\ref{eqn:diffR}). The axial propagation in state $w$ is given by $\frac{1}{2}[R_+(x,t)+R_-(x,t)]$ (Eq.~\ref{eqn:walkR}).

We construct a transition matrix of propagators $\mathbf{H}$, where $H_{a,b}(x,t)$ is the joint probability density for the time and position of a particle initially at the origin in state $a$ making its first transition out of that state, into state $b$. 
A Fourier transform in space $x\rightarrow k$ and a Laplace transform in time $t\rightarrow s$ is carried out to yield the transformed propagator $\hat{\widetilde{\mathbf{H}}}(k,s)$. The components of this propagator matrix are derived from the Laplace-transformed solutions for first passage times to an inner or outer absorbing boundary for a particle diffusing in a cylindrical domain~\cite{ozisik1985heat}. These components are given by,

\begin{equation}
\begin{split}
\fl{H}_{f_u,n} & = \frac{2\alpha}{(1-\alpha^2)\sigma_D}\frac{I_1(\sigma_D) K_1(\alpha \sigma_D) - I_1(\alpha\sigma_D) K_1(\sigma_D)}{I_0(\alpha\sigma_D) K_1( \sigma_D) + K_0(\alpha\sigma_D) I_1( \sigma_D)},\\
\fl{H}_{f,n} & = \frac{I_0((\alpha+\epsilon)\sigma_D) K_1( \sigma_D) + K_0((\alpha+\epsilon)\sigma_D) I_1( \sigma_D)}{I_0(\alpha\sigma_D) K_1( \sigma_D) + K_0(\alpha\sigma_D) I_1( \sigma_D)},\\
\fl{H}_{n_u,f} & = \frac{2}{\alpha\sigma_b} \frac{I_1(\alpha\sigma_b)}{I_0(\alpha \sigma_b)}, \quad 
\fl{H}_{n,f}  =  \frac{I_0((\alpha-\epsilon)\sigma_b)}{I_0(\alpha \sigma_b)} \\
\fl{H}_{n_u,w} & = \frac{\gamma}{D\sigma_b^2} \left(1 - \fl{H}_{n_u,f}\right), \quad
\fl{H}_{n,w}  = \frac{\gamma}{D\sigma_b^2} \left(1 - \fl{H}_{n,f}\right) \\
\fl{H}_{n_u,h} & = \frac{k_b}{D\sigma_b^2} \left(1 - \fl{H}_{n_u,f}\right),
 \quad
\fl{H}_{n,h}  = \frac{k_b}{D\sigma_b^2} \left(1 - \fl{H}_{n,f}\right) \\
\fl{H}_{h,w} & = \frac{\gamma}{s+\gamma+k_u}, \quad \fl{H}_{h,n_u} = \frac{k_u}{s+\gamma+k_u} \\
\fl{H}_{w,n_u} & = \frac{\lambda (s+\lambda)}{(s+\lambda)^2 + k^2},
\end{split}
\label{eqn:matrixcomponents}
\end{equation}

where  $\sigma_D = \sqrt{(s+Dk^2)/D}$, $\sigma_b = \sqrt{(s+\gamma+k_b+Dk^2)/D}$ and $I_\nu,\, K_\nu$ are the modified Bessel functions of order $\nu$ of the first and second kind, respectively. All other components of $\fl{\mathbf{H}}$ not listed in Eq.~\ref{eqn:matrixcomponents} correspond to transitions not allowed in the model and are equal to 0. To calculate the overall distribution of particles, we additionally define a vector of propagators $\fl{\mathbf{F}}$. Each component $\fl{F}_a$ corresponds to the Fourier-Laplace transformed spatial distribution of particles that first reached state $a$ at time $0$ and have moved a displacement $x$ at time $t$, without having left that state. These components are given by
\begin{equation}
\begin{split}
\fl{F}_{f_u} & = \frac{1}{D\sigma_D^2} \( 1 - \fl{H}_{f_u,n}\), \quad
\fl{F}_{f}  = \frac{1}{D\sigma_D^2} \( 1 - \fl{H}_{f,n}\)\\
\fl{F}_{n_u} & = \frac{1}{D\sigma_b^2} \( 1 - \fl{H}_{n_u,f}\), \quad
\fl{F}_{n}  = \frac{1}{D\sigma_b^2} \( 1 - \fl{H}_{n,f}\) \\
\fl{F}_{w} & = \frac{s+\lambda}{(s+\lambda)^2 + k^2}, \quad
\fl{F}_{h}  = \frac{1}{s+\gamma+k_u} .
\end{split}
\end{equation}

The overall propagator for a particle moving through this system of states can be found by a convolution over all possible transition paths, analogous to the discrete path sampling technique used for calculating kinetics on potential energy surfaces~\cite{wales2002discrete}. Specifically, the spatial density of a particle that started at the origin in state $i$ at time 0 and is in state $j$ at time $t$ is given by
\begin{equation}
\begin{split}
& G_{i,j}(x,t) =  \\
= &\delta_{i,j} F_i+ \sum_{n=0}^\infty \sum_{k_1,k_2,...k_n} H_{i,k_1} * \ldots * H_{k_{n-1},k_n} * H_{k_n,j} * F_j
\end{split}
\end{equation}
where $n$ is the number of intermediate states over which the particle transitions and $k_l$ is the identity of the $l$-th intermediate state.

Replacing the convolutions with multiplication of the Fourier-Laplace transformed propagators we find the overall spatial distribution for a particle that started in a linear combination of initial states described by the vector $\mathbf{P}$.
\begin{equation}
\begin{split}
\fl{G}(k,s; \mathbf{P}) = \lim_{\epsilon\rightarrow 0}\mathbf{P} \cdot \(\mathbf{I} - \fl{\mathbf{H}}\)^{-1} \cdot \fl{\mathbf{F}}
\end{split}
\label{eqn:tetherG}
\end{equation}
where $\mathbf{I}$ is the identity matrix.

The Laplace-transformed mean squared displacement can be found directly from the propagator by taking derivatives with respect to $k$. Its long time limit is found by expanding to lowest order in $s$ and taking the coefficient of the $1/s^2$ term:

\begin{equation}
\begin{split}
\lim_{t\rightarrow 0} \text{MSD} & = \left[-\lim_{s\rightarrow 0} s^2\( \left.\frac{\partial^2}{\partial k^2} \fl{G}(k,s)\right|_{k=0}\)\right]t = 2D_\text{eff} t, 
\end{split}
\end{equation}
where the effective long time diffusivity $D_\text{eff}$ is given in Eq.~\ref{eqn:defftether}.

\begin{figure}
	\centering
	\includegraphics[width=0.35\textwidth]{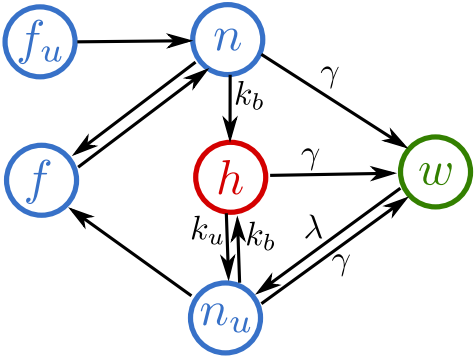}
	\caption{\label{fig:tetherstates}
		Schematic state diagram illustrating the particle states used to develop the analytical model for multi-modal transport in a cylinder. 
		Allowed transitions are labeled with arrows and the rates for the constant-rate transition processes (to and from tethered or actively walking state) are indicated. The transitions between diffusive states occur with a time-varying rate that can be derived by evaluating the matrix components in Eq.~\ref{eqn:matrixcomponents} at $k=0$.
		} 
	\label{fig:tetherstates}
\end{figure}

The average time for a particle with initial distribution $\mathbf{P}$ among the different states to first initiate a walk can be found 
as the time integral of the probability that no walk has yet occurred:
\begin{equation}
	\begin{split}
	\left<t_w(\mathbf{P})\right> = \int_0^\infty dt \int_0^\infty dx\; G^*(x,t; \mathbf{P}) 
	\end{split}
\end{equation}
where $G^*$ is obtained from Eq.\ref{eqn:tetherG} with alternate transition matrices $\fl{\mathbf{H}^*}, \fl{\mathbf{F}^*}$ defined by removing the rows and columns of  $\fl{\mathbf{H}}, \fl{\mathbf{F}}$ corresponding to the walking state (w).
The average time to start walking for a particle initially uniformly distributed within the inner radius $\alpha$ can be evaluated as
\begin{equation}
	\begin{split}
	\left<t_{w}\right> =  \left. \left[\(\mathbf{I} - \fl{\mathbf{H}^*}\)^{-1}\right]_{n_u,.} \cdot \fl{\mathbf{F}^*} \right|_{k=0,s=0}
	\end{split}
\end{equation}
where the subscript $(n_u,.)$ indicates the corresponding row of the inverse matrix. The resulting expression is given in Eq.~\ref{eqn:tstartw}.

\begin{figure}[ht!]
	\centering
	\includegraphics[width=\columnwidth]{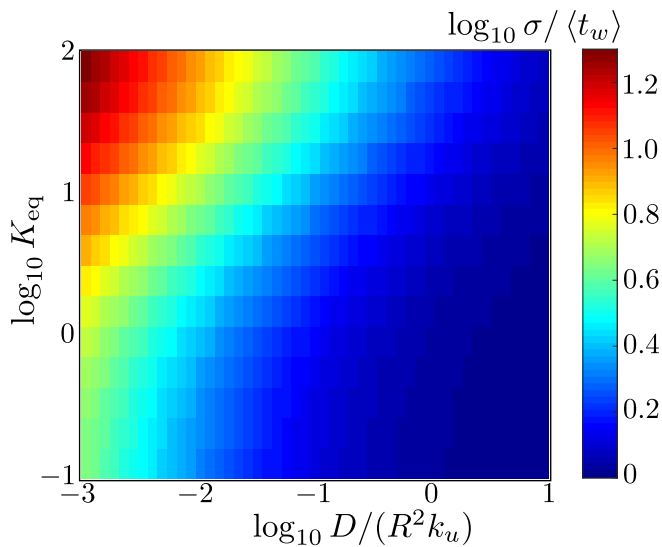}
	\caption{Fano factor $\sigma / \left<t_w\right>$ quantifying the variability in the time required for a particle to first begin a processive walk. The particle is assumed to start uniformly distributed within the inner radius $\alpha$. Results shown are for parameters $\alpha=0.1, k_u = 100, \gamma=10^{-2}$}
	\label{fig:fanowalk}
\end{figure}

We similarly calculate the mean squared time to initiate a walk, using
\begin{equation}
\begin{split}
\left<t_{w}^2\right> =  -2 \frac{\partial}{\partial s} \left. \left[\(\mathbf{I} - \fl{\mathbf{H}^*}\)^{-1}\right]_{n_u,.} \cdot \fl{\mathbf{F}^*} \right|_{k=0,s=0}.
\end{split}
\end{equation}
The variance in the time to start walking is given by $\sigma^2 = \left<t^2_w\right> - \left<t_w\right>^2$. While the full closed-form expression is too cumbersome to include here, in the limit of rapid unbinding from the tethered state ($k_u \gg \gamma, k_u\gg D/a^2$), the variance in the walking time is
\begin{equation}
\begin{split}
& \lim_{k_u\rightarrow \infty} \sigma^2 = \\ 
& \frac{4 D (1+\alpha^2K_\text{eq})^2 
	- \alpha^2\gamma(3-4\alpha^2+\alpha^4 + 4\log\alpha)(1+K_\text{eq}) }
{4\alpha^4 D \gamma^2 (1+K_\text{eq})^2}
\end{split}
\end{equation}
 Fig.~\ref{fig:fanowalk} shows the Fano factor, a measure of the variability in a stochastic process defined as the standard deviation in the time to start walking, relative to the average time. Large variability in how long it takes a passively diffusing particle to start walking is seen in the case of slow diffusion and strong binding.

\section{Simulation details}
\label{app:simdetails}

We simulate moving particles within a cylindrical domain of unit radius and unbounded length. The axial position of each particle is tracked to determine the range and the mean squared displacement. We also track the radial position to determine the probability of state transitions for the particles.

Each particle is assigned to a walking, diffusive or tethered state at initialization. The fraction of particles in each state is determined by the equilibrium distribution,
\begin{equation}
\begin{split}
f_{\mathrm{walk}} &= \frac{\gamma\alpha^2}{\gamma\alpha^2+\lambda\left(\frac{\alpha^2k_b+\gamma+k_u}{k_b+\gamma+k_u}\right)}\\
f_{\mathrm{diff}} &= \frac{\lambda\left(\frac{\gamma+k_u}{k_b+\gamma+k_u}\right)}{\gamma\alpha^2+\lambda\left(\frac{\alpha^2k_b+\gamma+k_u}{k_b+\gamma+k_u}\right)}\\
f_{\mathrm{bound}} &= \frac{\lambda\left(\frac{\alpha^2k_b}{k_b+\gamma+k_u}\right)}{\gamma\alpha^2+\lambda\left(\frac{\alpha^2k_b+\gamma+k_u}{k_b+\gamma+k_u}\right)}
\end{split}
\end{equation}
Unbound particles in the diffusive state start uniformly distributed radially throughout the cross section.

We divide the cylindrical domain in two concentric sections (Fig.~\ref{fig:tethering}a). The inner domain of radius $\alpha$ denotes the region within which particles can transition from the diffusive state to the walking or tethered state. Particles execute explicit Brownian dynamics with a time-step $\Delta t$ when their radial position is smaller than $3\alpha/2$. This includes the inner domain along with a buffer region of radius $\alpha/2$. The time step is chosen to be smaller than all relevant time-scales in the model:
 $\Delta t\ll \mathrm{min}(1/k_b,1/\gamma,\alpha^2/2D)$.  Note that this choice of time-step prevents multiple events occuring within a single step.

Particles outside the capture domain can spend a long time diffusing before reaching the region of interest. To accelerate the simulation, we make use of the first passage time distribution for diffusive particles between two cylindrical boundaries. The cumulative encounter probability to an absorbing inner boundary of radius $\alpha$ with a reflective outer boundary of unit radius is given by,
\begin{multline}
\Phi(t) = 1-\frac{\pi^2}{2}\sum_{n=1}^\infty \left[\frac{J_0^2(\beta_n \alpha)\beta_n\alpha}{J_0^2(\beta_n \alpha)-J_1^2(\beta_n)}\right]\\
\times\left[J_0(\beta_n r)Y_1(\beta_n)-Y_0(\beta_n r)J_1(\beta_n )\right]\\
\times\left[Y_1(\beta_n \alpha)J_1(\beta_n)-J_1(\beta_n \alpha)Y_1(\beta_n)\right]e^{-\beta_n^2Dt}
\end{multline}
where $J_\nu$ and $Y_\nu$ are the bessel functions of the first and second kind respectively, with order $\nu$~\cite{ozisik1985heat}. The $\beta_n$ are eigenvalues of the equation $J_1(\beta_n)Y_0(\beta_n\alpha)-J_0(\beta_n\alpha)Y_1(\beta_n) = 0$. The time required to reach the inner domain starting from an initial radial position $r$ is drawn from the above distribution and the particles are propagated along the axis according to the diffusive propagator $R_D$ (Eq.~\ref{eqn:diffR}) over this time interval.


The simulation is run using a hybrid Brownian Dynamics -- kinetic Monte Carlo algorithm where the probability of a state transition depends on the position of the particle. Particles in the diffusive state within the inner domain $(r <\alpha)$ can transition to the tethered or walking states at a combined rate $k_b+\gamma$. A transition is attempted at every diffusion time-step based on the relative probabilities for tethering and walking. Transitions leading away from the tethered state occur with a rate $k_u$ to the diffusive state, and with a rate $\gamma$ to the walking state. Particles in the walking state transition to the diffusive state a rate $\lambda$. Each time particles re-enter the diffusive state, they are uniformly distributed in the radial dimension within the inner region (of radius $\alpha$), ensuring symmetry between the binding and unbinding position. A schematic of these transitions is shown in Fig.~\ref{fig:tethering}(a).  For each transition out of a tethered or walking state, the waiting time is drawn from an exponential distribution with the mean equal to the corresponding transition rate. The particles are propagated in space according to the distribution for the given state over the duration of the waiting time. The simulation continues until all particles have covered a predetermined time interval. 

\bibliographystyle{aip} 
\bibliography{Bibliography} 
\end{document}